\documentclass[sigplan,10pt,review,anonymous]{article}
\usepackage{authblk}

\usepackage{geometry}

\usepackage{booktabs}   
\usepackage{subcaption} 

\usepackage{stmaryrd}
\usepackage{amsthm}

\newcounter{thmctr}

\newtheorem{theorem}{Theorem}[thmctr]

\newtheorem{lemma}[thmctr]{Lemma}
\newtheorem{example}[thmctr]{Example}
\newtheorem{definition}[thmctr]{Definition}
\newtheorem{proposition}[thmctr]{Proposition}

\usepackage{paralist}
\usepackage{bbm}
\usepackage{bbold}
\usepackage{listings}

\makeatletter
\newcommand{\xRightarrow}[2][]{\ext@arrow 0359\Rightarrowfill@{#1}{#2}}
\makeatother

\usepackage{varwidth}
\usepackage[inline]{enumitem}

\usepackage{amsmath}
\usepackage{amssymb} 
\usepackage{graphicx}
\usepackage{algorithm}
\usepackage{algorithmicx}
\usepackage[noend]{algpseudocode}
\usepackage{color}
\usepackage{xspace}
\usepackage{pgf}
\newcommand\pfun{\mathrel{\ooalign{\hfil$\mapstochar\mkern5mu$\hfil\cr$\to$\cr}}}
\newcommand{\dom}{\operatorname{\mathbf{dom}}}

\usepackage{tikz}
\usetikzlibrary{arrows,snakes,backgrounds,automata,shapes,positioning,chains,calc}

\usepackage[inference]{semantic}

\usepackage{marginfix} 

\newcommand{\OMIT}[1]{}

\usepackage{cutwin}

\begin{document}


\title{Verifying C11-Style Weak Memory Libraries}



\author{Sadegh Dalvandi and Brijesh Dongol
  \thanks{
    This work is supported by EPSRC Grant EP/R032556/1.
  }}

\affil{
  University of Surrey, Guildford, UK \\
  \texttt{\{m.dalvandi,b.dongol\}@surrey.ac.uk}}
\date{}


\newcommand{\linefill}{\cleaders\hbox{$\smash{\mkern-2mu\mathord-\mkern-2mu}$}\hfill\vphantom{\lower1pt\hbox{$\rightarrow$}}}  
\newcommand{\Linefill}{\cleaders\hbox{$\smash{\mkern-2mu\mathord=\mkern-2mu}$}\hfill\vphantom{\hbox{$\Rightarrow$}}}  
\newcommand{\transi}[2]{\mathrel{\lower1pt\hbox{$\mathrel-_{\vphantom{#2}}\mkern-8mu\stackrel{#1}{\linefill_{\vphantom{#2}}}\mkern-11mu\rightarrow_{#2}$}}}
\newcommand{\trans}[1]{\transi{#1}{{}}}
\newcommand{\eseq}[1]{\langle~\rangle}

\newcommand{\transo}[2]{\mathrel{\hbox{$\mathrel=_{\vphantom{#2}}\mkern-8mu\stackrel{#1}{\linefill_{\vphantom{#2}}}\mkern-11mu\Rightarrow_{#2}$}}}

\newcommand{\otrans}[1]{\transo{#1}{{}}}

\newcommand{\traceArrow}[1]{\stackrel{#1}{\longrightarrow}}
\newcommand{\ltsArrow}[1]{\stackrel{#1}{\Longrightarrow}}

\newcounter{sarrow}
\newcommand\strans[1]{%
  \mathrel{\raisebox{0.1em}{
    \stepcounter{sarrow}%
    \!\!\!\!
    \begin{tikzpicture}
      \node[inner sep=.5ex] (\thesarrow) {$\scriptstyle #1$};
      \path[draw,<-,decorate,line width=0.25mm,
      decoration={zigzag,amplitude=0.7pt,segment length=1.2mm,pre=lineto,pre length=4pt}] 
      (\thesarrow.south east) -- (\thesarrow.south west);
    \end{tikzpicture}%
  }}}

\newcommand\Strans[1]{%
\mathrel{\raisebox{0.1em}{
\!\!\begin{tikzpicture}
  \node[inner sep=0.6ex] (a) {$\scriptstyle #1$};
  \path[line width=0.2mm, draw,implies-,double distance between line
  centers=1.5pt,decorate, 
    decoration={zigzag,amplitude=0.7pt,segment length=1.2mm,pre=lineto,
    pre   length=4pt}] 
    (a.south east) -- (a.south west);
\end{tikzpicture}}%
}}

\newcommand{\calE}{{\cal E}}
\newcommand{\nat}{\mathbb{N}}
\newcommand{\noteq}{\neq}

\newcommand{\lt}{{\bf Less than}}

\newcommand{\ev}{\mathit{ev}}
\newcommand{\Events}{\mathit{Evt}}
\newcommand{\Inv}{\mathit{Inv}}
\newcommand{\Resp}{\mathit{Res}}
\newcommand{\his}{\mathit{his}}
\newcommand{\exec}{\mathit{exec}}
\newcommand{\complete}{\mathit{complete}}
\newcommand{\Var}{\mathit{Var}}
\newcommand{\GVar}{{\it GVar}}
\newcommand{\LVar}{\mathit{LVar}}
\newcommand{\Val}{\mathit{Val}}
\newcommand{\Obj}{\mathit{Obj}}
\newcommand{\Meth}{\mathit{Meth}}
\newcommand{\Tid}{\mathit{Tid}}
\newcommand{\CASOp}{\mathit{CAS}} 
\newcommand{\WR}{\mathsf{W_R}}
\newcommand{\RA}{\mathsf{R_A}}
\newcommand{\R}{\mathsf{R}}
\newcommand{\A}{\mathsf{A}}
\newcommand{\RX}{\mathsf{R_X}}
\newcommand{\W}{\mathsf{W}}
\newcommand{\WX}{\mathsf{W_X}}
\newcommand{\CRA}{\mathsf{CRA}}
\newcommand{\URA}{\mathsf{U}}
\newcommand{\URAT}{\mathsf{UT}}
\newcommand{\URAF}{\mathsf{UF}}

\newcommand{\HB}{{\sf hb}\xspace} 
\newcommand{\PO}{{\sf po}\xspace}
\newcommand{\MO}{{\sf mo}\xspace}
\newcommand{\SC}{{\sf sc}\xspace}
\newcommand{\RF}{{\sf rf}\xspace}
\newcommand{\SB}{{\sf sb}\xspace}

\newcommand{\refeq}[1]{(\ref{#1})}
\newcommand{\refalg}[1]{Algorithm~\ref{#1}}

\newcommand{\fr}{{\sf fr}}
\newcommand{\ltsb}{{\sf sb}}
\newcommand{\ltrf}{\mathord{\sf rf}}
\newcommand{\ltfr}{{\sf fr}}
\newcommand{\lthb}{{\sf hb}}
\newcommand{\ltsw}{{\sf sw}}
\newcommand{\ltmox}{{\sf mo}^x}
\newcommand{\ltmo}{{\sf mo}}
\newcommand{\lteco}{{\sf eco}}
\newcommand{\PreExec}{{\it PreExec}}
\newcommand{\Approx}{{\it C11}}
\newcommand{\Seq}{{\it Seq}}

\newcommand{\True}{{\it true}}
\newcommand{\False}{{\it false}}

\newcommand{\justified}{justified\xspace}
\newcommand{\notjustified}{unjustified\xspace}
\newcommand{\Justified}{Justified\xspace}
\newcommand{\Notjustified}{Unjustified\xspace}

\newcommand{\rdval}{{\it rdval}}
\newcommand{\wrval}{{\it wrval}}
\newcommand{\loc}{{\it loc}}
\newcommand{\var}{\mathtt{var}}

\newcommand{\imp}{\Rightarrow}
\newcommand{\expr}{\mathit{Exp}}
\newcommand{\flag}{\mathit{flag}}
\newcommand{\turn}{\mathit{turn}}
 
\newcommand{\hbo}[1]{\stackrel{#1}{\rightarrow}}
\newcommand{\detval}[1]{\stackrel{#1}{=}}
\newcommand{\last}{\mathit{last}}

\newcommand{\dview}{{\it dview}}
\newcommand{\wfs}{{\it wfs}}
\newcommand{\finite}{{\it finite}}

\newcommand{\kwcas}{\textsf{\textbf{CAS}}}
\newcommand{\kwswap}{\textsf{\textbf{swap}}}
\newcommand{\kwskip}{\bot}
\newcommand{\kwdo}{\textsf{\textbf{do}}}
\newcommand{\kwwhile}{\textsf{\textbf{while}}}
\newcommand{\kwend}{\textsf{\textbf{end}}}
\newcommand{\kwif}{\textsf{\textbf{if}}}
\newcommand{\kwthen}{\textsf{\textbf{then}}}
\newcommand{\kwelse}{\textsf{\textbf{else}}}
\newcommand{\kwreturn}{\textsf{\textbf{return}}}
\newcommand{\kwthread}{\textsf{\textbf{thread}}}
\newcommand{\kwuntil}{\textsf{\textbf{until}}}

\newcommand{\whilestep}[1]{\stackrel{#1}{\longrightarrow}}
\newcommand{\fv}{\mathit{fv}}

\algnewcommand\Swap{\kwswap}
\algnewcommand\Skip{\kwskip}
\algnewcommand\Thread{\kwthread}

\algrenewcommand\algorithmicend{\kwend}
\algrenewcommand\algorithmicdo{\kwdo}
\algrenewcommand\algorithmicwhile{\kwwhile}
\algrenewcommand\algorithmicfor{\textsf{\textbf{for}}}
\algrenewcommand\algorithmicforall{\textsf{\textbf{for all}}}
\algrenewcommand\algorithmicloop{\textsf{\textbf{loop}}}
\algrenewcommand\algorithmicrepeat{\textsf{\textbf{repeat}}}
\algrenewcommand\algorithmicuntil{\textsf{\textbf{until}}}
\algrenewcommand\algorithmicprocedure{\textsf{\textbf{procedure}}}
\algrenewcommand\algorithmicfunction{\textsf{\textbf{function}}}
\algrenewcommand\algorithmicif{\kwif}
\algrenewcommand\algorithmicthen{\kwthen}
\algrenewcommand\algorithmicelse{\kwelse}
\algrenewcommand\algorithmicreturn{\kwreturn}

\algblockdefx{Thread}{EndThread}%
[1]{\kwthread \xspace #1}%
{\algorithmicend}

\algblockdefx{MyWhile}{EndMyWhile}%
[1]{\kwwhile \xspace #1}%
{\algorithmicend}

\makeatletter
\ifthenelse{\equal{\ALG@noend}{t}}%
  {\algtext*{EndMyWhile}}
  {}%
\makeatother

\algblockdefx{MyUntil}{EndMyUntil}%
[1]{\kwuntil \xspace #1}%
{\algorithmicend}
\makeatletter
\ifthenelse{\equal{\ALG@noend}{t}}%
  {\algtext*{EndMyUntil}}
  {}%
\makeatother

\makeatletter
\ifthenelse{\equal{\ALG@noend}{t}}%
  {\algtext*{EndThread}}
  {}%
\makeatother

\newcommand{\action}[3]{\ensuremath{
\begin{array}[t]{l@{~}l}
\multicolumn{2}{l}{#1}\\
\textsf{Pre:}&#2\\
\textsf{Eff:}&#3
\end{array}
}}

\newcommand{\lstate}{\rho}
\newcommand{\llbr}[1]{\llbracket #1 \rrbracket}
\newcommand{\fresh}{\mathit{fresh}}

\newcommand{\kwtag}{{\it tag}}
\newcommand{\tid}{{\it tid}}
\newcommand{\act}{{\it act}}
\newcommand{\Op}{A}
\newcommand{\Prog}{{\it Prog}}
\newcommand{\Comm}{{\it Com}}
\newcommand{\AComm}{{\it ACom}}
\newcommand{\Exp}{{\it Exp}}
\newcommand{\CExp}{{\it CExp}}
\newcommand{\Init}{\mathbf{Init}}

\newcommand{\initq}{\mathit{qinit}}
\newcommand{\enq}{\mathit{enq}}
\newcommand{\deq}{\mathit{deq}}

\newcommand{\inits}{\mathit{sinit}}
\newcommand{\push}{\mathit{push}}
\newcommand{\pop}{\mathit{pop}}

\newcommand{\Sur}{{\it RC11}}
\newcommand{\asgn}{\ensuremath{:=}}

\newcommand{\bbD}{\mathbb{D}}
\newcommand{\bbE}{\mathbb{E}}
\newcommand{\bbS}{\mathbb{S}}

\newcommand{\rat}{\mathbb{Q}}
\newcommand{\bool}{\mathbb{B}}

\newcommand{\cons}{cons}

\newcommand{\matchedTS}{{\tt matched}}

\newcommand{\maxmo}{\mathbf{max}_{\ltmo}}
\newcommand{\cclose}{\mathbf{cclose}}
\newcommand{\scomp}{\circ}
\newcommand{\view}{\mathit{View}}
\newcommand{\tview}{{\tt tview}}
\newcommand{\ctview}{{\tt ctview}}
\newcommand{\writeson}{{\tt writes\_on}}
\newcommand{\ls}{\mathit{ls}}
\newcommand{\rdview}{\mathit{rdview}}
\newcommand{\isReleasing}{\mathit{isRel}}
\newcommand{\mview}{{\tt mview}}
\newcommand{\mods}{\mathtt{mods}}
\newcommand{\writes}{\mathtt{ops}}
\newcommand{\covered}{\mathtt{cvd}}
\newcommand{\enc}{\mathit{enc}}

\newcommand{\ts}{{\it ts}}

\newcommand{\tst}{{\tt tst}}
\newcommand{\OW}{\mathtt{Obs}}
\newcommand{\visWrites}{\mathit{V\!W}\!}
\newcommand{\encounteredWrites}{\mathit{E\!W}\!}
\newcommand{\Act}{{\sf Act}}

\newcommand{\acquire}{{\it acquire}}
\newcommand{\release}{{\it release}}
\newcommand{\init}{{\it init}}
\newcommand{\maxTS}{{\it maxTS}}

\newcommand{\eqrng}[2]{(\ref{#1}-\ref{#2})}
\newcommand{\refprop}[1]{Proposition~\ref{#1}}
\newcommand{\reffig}[1]{Figure~\ref{#1}}
\newcommand{\refthm}[1]{Theorem~\ref{#1}}
\newcommand{\reflem}[1]{Lem\-ma~\ref{#1}}
\newcommand{\refcor}[1]{Corollary~\ref{#1}}
\newcommand{\refsec}[1]{Section~\ref{#1}}
\newcommand{\refex}[1]{Example~\ref{#1}}
\newcommand{\refdef}[1]{Definition~\ref{#1}}
\newcommand{\reflst}[1]{Listing~\ref{#1}}
\newcommand{\refchap}[1]{Chapter~\ref{#1}}
\newcommand{\reftab}[1]{Table~\ref{#1}}

\newcommand{\kwfai}{\textsf{\textbf{FAI}}}
\newcommand{\vwrites}{{\tt visible\_writes}}
\newcommand{\wrts}{{\tt writes}}
\newcommand{\isCovered}{\mathit{isCovered}}

\newcommand{\WrX}[2]{#1 := #2}
\newcommand{\WrR}[2]{#1 :=^{\sf R} #2}
\newcommand{\RdX}[2]{#1 \gets #2}
\newcommand{\RdA}[2]{#1 \gets^{\sf A} #2}
\newcommand{\CObs}[5]{\langle #1 = #2\rangle [#4 = {#5}]_{#3}}
\newcommand{\CObss}[3]{\langle #1\rangle [#2]_{#3}}

\newcommand{\COSem}[2]{#1[#2]}
\newcommand{\cvd}[2]{{\bf C}_{#1}^{#2}}
\newcommand{\cvv}[2]{{\bf H}_{#1}^{#2}}

\tikzset{
    mo/.style={dashed,->,>=stealth,thick,black!20!purple},
    hb/.style={solid,->,>=stealth,thick,blue},
    sw/.style={solid,->,>=stealth,thick,black!50!green},
    rf/.style={dashed,->,>=stealth,thick,black!50!green},
    fr/.style={dashed,->,>=stealth,thick,red}
 }

 \lstset{
    mathescape=true,
    breaklines=false,
    tabsize=2,
    morekeywords={if, then, else, let, in },keywordstyle=\color{blue},
    basicstyle=\ttfamily,
    literate={\ \ }{{\ }}1
  }

\maketitle

  \begin{abstract}
    Deductive verification of concurrent programs under weak memory
    has thus far been limited to simple programs over a monolithic
    state space.
    For scalabiility, we also require 
    modular techniques with verifiable library abstractions. This
    paper addresses this challenge in the context of RC11 RAR, a
    subset of the C11 memory model that admits relaxed and
    release-acquire accesses, but disallows, so-called, load-buffering
    cycles. We develop a simple  framework for specifying
    abstract objects that precisely characterises the observability
    guarantees of abstract method calls. We show how this framework
    can be integrated with an operational semantics that enables
    verification of client programs that execute abstract method calls
    from a library it uses. Finally, we show how implementations of
    such abstractions in RC11 RAR can be verified by developing a
    (contextual) refinement framework for abstract objects. Our
    framework, including the operational semantics, verification
    technique for client-library programs, and simulation between
    abstract libraries and their implementations, has been mechanised
    in Isabelle/HOL.
  \end{abstract}





\section{Introduction}



An effective technique for reasoning about weak memory models is to
consider the observations that a thread can make of the
writes 
within a system.  For example, for certain subsets of C11 (the 2011 C
standard), 
reasoning about per-thread observations has led to operational
characterisations of the memory model, high-level predicates for
reasoning about per-thread observations, and deductive verification
techniques applied to standard litmus tests and synchronisation
algorithms~\cite{ECOOP20}.
Current verification techniques are however, focussed on (closed)
programs, and hence 
do not provide any mechanism for (de)composing clients
and libraries. This problem requires special consideration under weak
memory since the execution of a library
method 
induces synchronisation. That is, a thread's observations of a system
(including of client variables) can change 
when executing 
library methods.

This paper addresses several questions surrounding client-library
composition in a weak memory context.

{\bf (1)} {\em How can a client \emph{use} a weak memory library,
i.e., what abstract guarantees can a library provide a client program?}
Prior works~\cite{DongolJRA18,DBLP:conf/popl/BattyDG13} describe
techniques for \emph{specifying} the behaviour of abstract objects,
which are in turn related to their implementations using causal
relaxations of linearizability. 
However, these works do not provide a mechanism for reasoning about
the behaviour of client programs that {\em use} abstract libraries. In
this paper, we address this gap by presenting a modular operational
semantics 
that combines weak
memory states of clients and libraries.

{\bf (2)} {\em What does it mean to \emph{implement} an abstract
  library?}  To ensure that behaviours of client programs using an
abstract library are preserved, we require \emph{contextual
  refinement} between a library implementation and its abstract
specification. This guarantees that no new client behaviours are
introduced when a client uses a (concrete) library implementation in
place of its (abstract) library specification. Under sequential
consistency (SC), it is well known that linearizable libraries
guarantee (contextual)
refinement~\cite{DBLP:conf/icfem/DongolG16,GotsmanY11,DBLP:journals/tcs/FilipovicORY10}. However,
under weak memory, a generic notion of linearizability is difficult to
pin down~\cite{DongolJRA18,ifm18}. We therefore present a direct
technique for establishing contextual refinement under weak memory. A
key innovation is the development of context-sensitive simulation
rules that ensures that each client thread that uses the
implementation observes a subset of the values seen by the
abstraction. 


{\bf (3)} {\em Can the same abstract library specify
  \emph{multiple} implementations?}  A key benefit of refinement is
the ability to use the same abstract specification for multiple
implementations, e.g., to fine-tune clients for different concurrent
workload scenarios. To demonstrate applicability of our framework, we
provide a proof-of-concept example for an abstract lock and show that
the same lock specification can be implemented by a sequence lock and
ticket lock. The theory itself is generic and can be applied to
concurrent objects in general.

{\bf (4)} {\em How can we support verification? Can the
  verification techniques be mechanised?}
Assuming the existence of an operational semantics for the underlying
memory model, we aim for \emph{deductive} verification of both
client-library composition and contextual refinement. We show that
this can be supported 
by prototyping the full
verification stack in the Isabelle/HOL theorem prover~\footnote{Our
  Isabelle theories may be accessed as ancillary material in the ArXiV submission.}.  




\section{Message passing via library objects}
\label{sec:message-passing-via}



In this section, we illustrate the basic principles of client-object synchronisation in weak memory. 

\begin{figure}[t]
  \begin{minipage}[b]{0.45\columnwidth}
      \begin{center} 
  {\bf Init: } $d:=0;$ $s.init();$  \\
  $\begin{array}{@{}l@{\ }||@{\ }l}
     \text{\bf Thread } 1
     & \text{\bf Thread } 2\\
     d := 5; \qquad & 
                       \text{\bf do } r_1 := s.pop() \\
          
     s.push(1); & \text{\bf until}\ r_1 = 1;  \\ 
     & r_2 \gets d; \\
     \end{array}$

   {\color{red!70!black} $\{r_2 = 0 \lor r_2=5\}$}  \qquad \quad  \quad     
 \end{center}
 \vspace{-1em}
 \caption{Unsynchronised message passing}
 \label{fig:po-message-bad}
\end{minipage}
\hfill
  \begin{minipage}[b]{0.47\columnwidth}
  \begin{center}  
  {\bf Init:} $d:=0;$ $s.init();$ \\
  $\begin{array}{@{}l@{\ }||@{\ }l}
     \text{\bf Thread } 1
     & \text{\bf Thread } 2\\
     d := 5; \qquad & 
                      \text{\bf do } r_1 := s.pop^{\sf A}() \\
     s.push^{\sf R}(1);               & \text{\bf until}\ r_1 = 1;  \\ 
     
     & r_2 \gets d; \\
 
     \end{array}$

   {\color{green!40!black} $\{ r_2=5\}$}  \qquad   \  \ \ 
 \end{center}
  \vspace{-1em}
 \caption{Publication via a synchronising
   stack 
 }
 \label{fig:publication}
\end{minipage}

\vspace{-1.3em}
 \end{figure}

\smallskip \noindent{\bf Client-object message passing.}
Under SC all threads have a single common view of the shared
state. When a new write is executed, the ``views'' of all threads are
updated so that they are guaranteed to only see this new write. In
contrast, each thread in a C11 program has its own view of each
variable. Views may not be updated when a write occurs, allowing
threads to read stale writes. To enforce view updates, additional
synchronisation (e.g., release-acquire) must be introduced~\cite{DBLP:conf/popl/BattyOSSW11,DBLP:conf/ecoop/KaiserDDLV17,DBLP:conf/pldi/LahavVKHD17}.


Now consider a generalisation of this idea to (client) programs that use
library
objects. 
The essence of the problem is illustrated by the message-passing programs in Figures~\ref{fig:po-message-bad}~and~\ref{fig:publication}. 
Under SC, when the program in
\reffig{fig:po-message-bad} terminates, the value of $r_2$ is
guaranteed to be $5$. 
However, this is not necessarily true in a weak memory setting. Even
if $pop$ operation in thread~2 returns 1, it may be possible for
thread 2 to observe stale value 0 for $d$. Therefore the program only
guarantees the weaker postcondition $r_2 = 0 \lor r_2 = 5$.

To address this problem, the library operations in
\reffig{fig:publication} are annotated with release-acquire
annotations. In particular, the client assumes the availability of a
``releasing push'' ($push^{\sf R}(1)$), which is to be used for
message passing. Thread~2 pops from $s$ using an ``acquiring pop''
($pop^{\sf A}()$). If this pop returns 1, the stack operations induce
a happens-before synchronisation in the client, which in turn means
that it is now impossible for thread 2 to read the stale initial
write for $d$.




\smallskip\noindent{\bf Verification strategy.}
Our aim is to enable {\em deductive verification} of such programs by
leveraging recently developed operational semantics, assertion
language and Owicki-Gries style proof strategy for RC11
RAR~\cite{ECOOP20}. We show that these existing concepts generalise
naturally to client-object, and in a manner that enables modular
proofs.

The assertion language of~\cite{ECOOP20} enables reasoning about a
thread's views, e.g., in \reffig{fig:mp-proof}, after initialisation,
thread $t \in \{1,2\}$ has \textit{definite value} $0$ for $d$
(denoted $[d = 0]_t$).

In this paper, we extend such assertions to capture thread views over
library objects. E.g., after initialisation, the only value a pop by
thread $t$ can return is $Empty$, and this is captured by the
assertion $[s.pop_{emp}]_t$. 
The precondition of $d := 5$ states that thread 2 cannot pop value
$1$ from $s$ (as captured by the assertion $\neg \langle
s.pop_1\rangle_2$).
The precondition of the $\kwuntil$ loop in thread 2 contains a {\em
  conditional observation} assertion (i.e.,
$\langle s.pop_1 \rangle [d = 5]_2$), which states that if thread~2
pops value 1 from $s$ then it will subsequently be in a state where it
will definitely read $5$ for $d$.

A key benefit of the logic in~\cite{ECOOP20} is that it enables use of
{\em standard} Owicki-Gries reasoning and straightforward
mechanisation~\cite{DBLP:journals/corr/abs-2004-02983}. As we shall see (\refsec{sec:example-client-lbjec}), we maintain these benefits in
the context of client-object programs.



\begin{figure}[t]
  \centering
 

\begin{minipage}[t]{0.9\columnwidth}
    \begin{center} 
  {\bf Init: } $d:=0;$ $s.init();$  \qquad \qquad \qquad \qquad\\ 
   {\color{green!40!black} $\{[d = 0]_1 \wedge [d = 0]_2 \wedge [s.pop_{emp}]_1 \wedge [s.pop_{emp}]_2\}$} \\ 
  $\begin{array}{l@{\quad}||@{\quad }l}
     \text{\bf Thread } 1
     & \text{\bf Thread } 2\\
     \begin{array}[t]{@{}l@{}}
       {\color{blue} \{ \neg \langle s.pop_1 \rangle_2 \wedge [d = 0]_1\}} \\
       1: d := 5; \\
       {\color{blue} \{ \neg \langle s.pop_1 \rangle_2 \wedge [d = 5]_1\}} \\
       2: s.push^{\sf R}(1); \\
       {\color{blue} \{ true\}}
     \end{array}
     & 
       \begin{array}[t]{@{}l@{}}
         {\color{blue} \{ \langle s.pop_1 \rangle [d= 5]_2 \}} \\ 
         3: \text{\bf do } r_1 := s.pop^{\sf A}()\ \\
         \text{\bf   until}\ r_1 = 1; \\
         {\color{blue} \{ [d = 5]_2\}} \\
         4: r_2 \gets d; \\
         {\color{blue} \{r_2 = 5\} }
     \end{array}
   \end{array}
   $

   {\color{green!40!black} $\{ r_2=5\}$} 
 \end{center}
 \vspace{-1em}
  \caption{A proof outline for message passing}
  \label{fig:mp-proof}
\end{minipage}

\end{figure}

\smallskip\noindent{\bf Contextual refinement.}
\label{sec:cont-refin-pre}
Contextual refinement relates a client using an abstract object with a
client that uses a concurrent implementation of the object. More precisely, we say
that a concrete object $CO$ is a \emph{contextual refinement} of an
abstract object $AO$ iff for any client $C$, every behaviour of $C$
when it uses $CO$ is a possible behaviour of $C$ when it uses
$AO$. Thus, there is no observable difference to any client when it
uses $CO$ in place of $AO$.

In a weak memory setting, to enable a client to \emph{use} an object,
one must \emph{specify} how synchronisation between object method
calls affects the client state. To {\em implement} such a
specification, we must describe how the abstract synchronisation
guarantees 
are represented in the implementation. Prior works have appealed to
extensions of notions such as linearizability to ensure contextual
guarantees
\cite{DongolJRA18,DBLP:journals/pacmpl/EmmiE19,DBLP:journals/pacmpl/RaadDRLV19}. In
this paper, we aim for a more direct approach and consider contextual
refinement 
 directly. 



\section{Generalised 
  operational semantics}
\label{sec:gener-weak-memory}
We now present a simple program syntax 
that allows one to write open programs 
that can be filled by an abstract method or concrete implementation of
a method.


\subsection{Program Syntax}
\label{sec:program-syntax}

We start by defining a syntax of concurrent programs, starting with
the structure of sequential programs (single threads).  A thread may
use {\em global} shared variables (from $\GVar$) and local registers
(from $\LVar$).  We let $\Var = \GVar \cup \LVar$ and assume
$\GVar \cap \LVar = \emptyset$. For client-library programs, we 
partition $\GVar$ into $\GVar_C$ (the global client variables) and
$\GVar_L$ (the global library variables) and similarly $\LVar$ into
$\LVar_C$ and $\LVar_L$. In an implementation, global variables can be
accessed in three different {\em synchronisation modes}: acquire ({\sf
  A}, for reads), release ({\sf R}, for writes) and relaxed (no
annotation).  The annotation {\sf RA} is employed for {\em update}
operations, which reads and writes to a shared variable in a single
atomic step.  
We let $\Obj$ and $\Meth$ be the set of all objects and method calls,
respectively.

We assume that
$\ominus$ is a unary operator (e.g., $\neg$), $\oplus$ is a binary
operator (e.g., $\land$, $+$, $=$) and $n$ is a value (of type
$\Val$). Expressions must only involve local
variables. 
The syntax of sequential programs, $\Comm$, is given by the following
grammar with $r \in \LVar, x \in \GVar, o \in \Obj, m \in \Meth, u, v \in \Val$:

\medskip
\noindent
\begin{tabular}[t]{r@{~}l}
  $\Exp_L$   ::= & $\Val \mid \LVar  \mid \ominus \Exp_L \mid \Exp_L \oplus Exp_L$ \\[1mm]
  $\CExp_L$   ::= & $\bullet \mid \Exp_L$ \\[1mm]
  $\bullet$ ::= & $\Val \mid o.m([u]) \mid \Comm $, where $\Comm$ contains no holes \\[1mm]
  
  $\AComm$ ::= & $ \bullet \mid \kwskip \mid 
                 r \gets \kwcas(x, u, v)^{\sf RA} 
\mid 
                 r \gets \kwfai(x)^{\sf RA}  
                   \mid r := \CExp_L  
                   \mid x :=^{\sf [R]} \Exp_L \mid  r \gets^{\sf [A]} x$ \\[6pt] 
  $\Comm$ ::= & $\AComm \mid \Comm ; \Comm \mid \kwif~B\ \kwthen\ \Comm\ \kwelse\ \Comm \mid  
                    \kwwhile\ B\ \kwdo\ \Comm$
\end{tabular} \medskip

\noindent where we assume $B$ to be an expression of type $\CExp_L$
that evaluates to a boolean. 
We allow programs with holes, denoted
$\bullet$, which may be filled by an abstract or concrete method
call. During a program's execution, the hole may also be filled by
the null value $\bot \notin
\Val$, or the return value of the method call.
The notation ${\sf [X]}$ denotes that the annotation ${\sf X}$ is
optional, where ${\sf X} \in \{{\sf A}, {\sf R}\}$, enabling one to
distinguish relaxed, acquiring and releasing accesses. Within a method
call, the argument $u$ is optional. Later, we will also use
$\kwdo$-$\kwuntil$ loops, which is straightforward to define in terms
of the  syntax above.

\subsection{Program Semantics}

\label{sec:program-semantics}

For simplicity, we assume concurrency at the top level only. We let
$\Tid$ to be the set of all thread identifiers and use a function
$\Prog: \Tid \to \Comm$ to model a program comprising multiple
threads. In examples, we typically write concurrent programs as
$C_1 || \ldots || C_n$, where $C_i \in \Comm$.  We further assume some
initialisation of variables.  The structure of our programs
thus is $\Init; \big( C_1 || \ldots || C_n \big) $.

The operational semantics for this language is defined in three parts.
The \emph{program semantics} fixes the steps that the concurrent
program can take.  This gives rise to transitions
$(P,\lstate) \trans {a}_t (P',\lstate')$ of a thread $t$ where $P$ and
$P'$ are programs, $\lstate$ and $\lstate'$ is the state of local
variables and $a$ is an action (possibly the silent action $\epsilon$,
see below).  The program semantics is combined with a {\em memory
  semantics} which reflects the C11 state, and in particular the write
actions from which a read action can
read. 
Finally, there is the \emph{object semantics}, which defines the
abstract semantics of the object at hand.

We assume that the set of actions is given by $\Act$. We let
$\epsilon \notin \Act$ be a silent action and let
$\Act_\epsilon = \Act \cup \{\epsilon\}$.


In the program semantics, we assume a function
$\lstate \in \Tid \rightarrow (\LVar \pfun \Val)$, which returns the
local state for the given thread.  We assume that the local variables
of threads are disjoint, i.e., if $t \noteq t'$, then
$\dom(\lstate(t)) \cap \dom(\lstate(t')) = \emptyset$. For an
expression $E$ over local variables, we write $\llbr{E}_{\ls}$ for the
value of $E$ in local state $\ls$; we write $\ls[r := v]$ to state
that $\ls$ remains unchanged except for the value of local variable
$r$ which becomes $v$.

\begin{figure*}[t]\small
  \centering %
  $\inference{r \in \LVar \quad v = \llbr{E}_{{\it ls}}}{(r := E,{\it ls}) \trans{\epsilon} (\kwskip,\ls[r := v]) }
    \qquad
    \inference{x \in \GVar \quad a = wr^{\sf [R]}(x, \llbr{E}_{{\it ls}}) }{(x :=^{\sf [R]} E,ls) \trans{a} (\kwskip,\ls) }$ \bigskip
   
  $\inference{a = rd^{\sf [A]}(x,v) \quad v \in \Val}{ (r \gets^{\sf [A]} x, \ls) \trans a (\kwskip,\ls[r := v]) }
  $
   \bigskip
  
$
   \inference{(C_1,\ls) \trans{a} (C_1',ls')}{(C_1 ; C_2,\ls) \trans{a}
    (C_1' ; C_2,\ls')}
  \qquad
     \inference{v \in \Val \cup \{\bot\}}{(v ; C_2,\ls) \trans{\epsilon}
    (C_2,\ls)}$ \bigskip

  $
    \inference
  {\llbr{B}_{\ls}}
  {({\it IF}, \ls) \trans{\epsilon} (C_1,\ls)} \quad 
    \inference
  {\neg \llbr{B}_{\ls}}
  {({\it IF}, \ls) \trans{\epsilon} (C_2,\ls) }
  $
 \bigskip

 $
    \inference{\llbr{B}_{\ls}}{
    \begin{array}[t]{@{}l@{}}
      ({\it WHILE}, \ls) 
      \trans{\epsilon}  (C; {\it WHILE}, \ls) 
    \end{array}
  } \qquad 
  \inference{\neg \llbr{B}_{\ls}}
  {
    \begin{array}[t]{@{}l@{}}
      ({\it WHILE}, \ls)  \trans{\epsilon}  (\kwskip, \ls) 
    \end{array}
  }
  $\bigskip

  $
    \inference{a = rd(x,v') \quad v'\neq u \quad u,v,v' \in \Val}{ (r \gets \kwcas(x, u, v), \ls) \trans a (\kwskip,\ls[r := \False]) }
$
\bigskip

  $
    \inference{a = upd^{\sf RA}(x,u,v) \quad u,v \in \Val}{ (r \gets \kwcas(x, u, v), \ls) \trans a (\kwskip,\ls[r := \True]) }
    \qquad
    \inference{a = upd^{\sf RA}(x,u,u+1) \quad u \in \Val}{ (r \gets \kwfai(x), \ls) \trans a (\kwskip,\ls[r := u]) }
  $ \bigskip

  $
    \inference{}{ (\COSem{C}{\kwskip}, \ls) \trans \epsilon (C,\ls) }
  \qquad
  \inference{(D, ls) \trans a (D', ls')}{ (\COSem{C}{D}, \ls) \trans{a}_L (\COSem{C}{D'},\ls') }
  $\bigskip

$
  \inference[\sc
  Cli]{(P(t),\lstate(t)) \trans{a} (C,\ls) \quad a \in \Act_\epsilon}
  {(P,\lstate) \trans{a}_t (P[t := C],\lstate[t := \ls])}
  \qquad
  \inference[\sc
  Lib]{(P(t),\lstate(t)) \trans{a}_L (C,\ls) \quad a \in \Act_\epsilon}
  {(P,\lstate) \trans{a}_{L, t} (P[t := C],\lstate[t := \ls])}
$
\caption{Program semantics, where ${\it IF} = \kwif \ B\ \kwthen\
    C_1\ \kwelse\ C_2$ and ${\it WHILE} = \kwwhile\ B\ \kwdo\ C$}
  \label{fig:comm-sem}
\end{figure*}

We use $C[D]$ to denote the program $C$ with the leftmost innermost
hole filled by $D$. If $D = \bot$, we proceed with the execution of
$C$, otherwise we execute $D$. Note that if $D$ terminates with a
value (due to a method call that returns a value), then the hole
contains a value and execution may proceed by either using the rule
for $r \asgn v$ or the rule for $v ; C_2$, both of which are present
in \reffig{fig:comm-sem}. The last two rules, {\sc Cli} and {\sc Lib},
lift the transitions of threads to a transition of a client and
library program, respectively. These are distinguished by the
subscript $L$, which only appears in transitions corresponding to the
library.

The rules in \reffig{fig:comm-sem} allow for {\em all} possible values
for any read. We constrain these values 
with respect to a {\em memory semantics} (formalised by
$\strans{a}_t$), which is described for reads, writes and updates in
\refsec{sec:memory-semantics} and for abstract objects in
\refsec{sec:abstr-object-semant}.
The combined semantics brings together a client state $\gamma$ and library
state $\beta$ as follows.
\begin{gather*} \small
  \inference{(P,\lstate) \trans {\epsilon }_t (P',\lstate')}
  {(P,\lstate,\gamma, \beta) \ltsArrow{
    } (P', \lstate',\gamma, \beta)}  \small
  \ \ \ 
  \inference{(P,\lstate) \trans {\epsilon }_{L,t} (P',\lstate')}
  {(P,\lstate,\gamma, \beta) \ltsArrow{
    }
    (P', \lstate',\gamma, \beta)}
\\[5pt] \small
  \inference{(P,\lstate) \trans {a}_t (P',\lstate') \\ \gamma,
    \beta \strans{a}_{t} \gamma', \beta'} {(P,\lstate,\gamma, \beta)
    \ltsArrow{
    } (P', \lstate',\gamma', \beta')} 
  \hfill
  \inference{(P,\lstate) \trans {a}_{L,t} (P',\lstate') \\
    \beta, \gamma \strans{a}_{t} \beta', \gamma'}
  {(P,\lstate,\gamma, \beta)
    \ltsArrow{
    } (P', \lstate',\gamma', \beta')}
\end{gather*}
These rules ensure, for example, that a read only returns a value
allowed by the underlying memory model. In
\refsec{sec:abstr-object-semant}, we introduce additional rules so
that the memory model also contains actions corresponding to method
calls on an abstract object.

Note that the memory semantics (see \refsec{sec:memory-semantics} and \refsec{sec:abstr-object-semant}) defined
by $\gamma, \beta \strans{a}_t \gamma', \beta'$ assumes that $\gamma$ is
the state of the component being executed and $\beta$ is the state of
the context. For a client step, we have that $\gamma$ is the
executing component state and $\beta$ is the context state, where as
for a library step, these parameters are swapped.

\subsection{Memory Semantics}
\label{sec:memory-semantics}

Next, we detail the modularised memory semantics, which builds on an
earlier monolithic semantics~\cite{ECOOP20}, which is a
timestamp-based revision of an earlier operational
semantics~\cite{DBLP:conf/ppopp/DohertyDWD19}. Our present extension
is a semantics that copes with client-library interactions in weak
memory. Namely, it describes how synchronisation (in our example
release-acquire synchronisation) in one component affects thread views
in another component. The semantics 
accommodates both client synchronisation affecting a library, and vice
versa.

\smallskip\noindent {\bf Component State.} 
We assume $\Act$ denotes the set of actions. Following~\cite{ECOOP20},
each global write is represented by a pair
$(a, q) \in \Act \times \rat$, where $a$ is a write action, and $q$ is
a rational number that we use as a {\em timestamp} corresponding to
modification order (cf.
\cite{DBLP:conf/ecoop/KaiserDDLV17,Dolan:2018:LDRF,DBLP:journals/corr/PodkopaevSN16}).
The set of modifying operations within a component that have occurred
so far is recorded in $\writes \subseteq \Act \times \rat$. Unlike
prior works, to accommodate (abstract) method
calls of a data structures, we record abstract operations in general,
as opposed to writes only.

Each state must record the operations that are observable to each
thread. To achieve this, we use two families of functions from global
variables to writes~(cf.
\cite{DBLP:journals/corr/PodkopaevSN16,DBLP:conf/popl/KangHLVD17}). 
\begin{itemize}[leftmargin=*]
\item A \emph{thread view} function $\tview_t \in \GVar \rightarrow \writes$ that returns the
  \emph{viewfront} of thread $t$. The thread $t$ can read from any
  write to variable $x$ whose timestamp is not earlier than
  $\tview_t(x)$. Accordingly, we define, for each state $\gamma$,
  thread $t$ and global variable $x$, the set of {\em observable
    writes}, where $\tst(w) = q$ denotes $w$'s timestamp:

  \smallskip\hfill$\gamma.\OW(t, x) =  \{(a, q) \in \gamma.\writes  \mid 
                                     \begin{array}[t]{@{}l@{}}
\mathit{var}(a) = x  \\  {}\wedge \tst(\gamma.\tview_t(x)) \leq q\}
                                     \end{array}$\hfill{}

\item A \emph{modification view} function $\mview_w \in  \GVar \rightarrow \Act \times \rat$ that records the
  \emph{viewfront} of write $w$, i.e., the viewfront of the thread
  that executed $w$ immediately after $w$'s execution. We use
  $\mview_w$ to compute a new value for $\tview_t$ if a thread $t$
  \emph{synchronizes} with $w$, i.e., if $w \in \WR$ and another
  thread executes an $e \in \RA$ that reads from $w$.
\end{itemize}
The client cannot directly access writes in the library, therefore the thread
view function must map to writes within the same component. On the other
hand, synchronisation in a component can affect thread views in
another (as discussed in \refsec{sec:message-passing-via}), thus the
modification view function may map to operations across the system.


Finally, our semantics maintains a set $\covered \subseteq
\writes$. In C11 RAR, each update action occurs in modification order
immediately after the write that it reads from
\cite{DBLP:conf/ppopp/DohertyDWD19}. This property ensures the
atomicity of updates. 
We disallow 
any newer modifying operation (write or update) from intervening
between any update and the write or update that it reads from.  As we
explain below, covered writes are those that are immediately prior to
an update in modification order, and new write actions never interact
with a covered write.

\smallskip
\noindent{\bf Initialisation.} 
Suppose
$\GVar_C = \{x_1, \ldots, x_n\}$, $\GVar_L = \{y_1, \ldots, y_n\}$, 
$\LVar = \{r_1, \ldots, r_m\}$,
$k_1, \dots, k_n, l_1, \dots, l_m \in \Val$, and
$\Init = x_1:=k_1; \ldots, x_n:=k_n ; [r_1 := l_1 ;] \dots [r_m :=
l_m;]$, where we use the notation $[r_i := l_i;]$ to mean that the
assignment $r_i := l_i$ may optionally appear in $\Init$. Thus each
shared variable is initialised exactly once and each local variable is
initialised at most once. The initial values of the state components
are then as follows, where we assume $0$ is the initial
timestamp, $t$ is a thread, $x_i \in \GVar_C$ and $y_i \in \GVar_L$
\begin{align*}
  \gamma_\Init.\writes & = \{(wr(x_1,k_1),0), \ldots, (wr(x_n,k_n),0)\} \\
  \beta_\Init.\writes & = \{(wr(y_1,k_1),0), \ldots, (wr(y_n,k_n),0)\} \\
  \gamma_\Init.\tview_t(x_i) & =  (wr(x_i,k_i),0)\\ 
  \beta_\Init.\tview_t(y_i) & =  (wr(y_i,k_i),0) \\
  \gamma_\Init.\mview_{x_i} &=\beta_\Init.\mview_{y_i}= \gamma_\Init.\tview_t \!\cup\! \beta_\Init.\tview_t \\
  \gamma_\Init.\covered &= \beta_\Init.\covered= \emptyset
\end{align*}

The local state component of each thread must also be compatible with
$\Init$, i.e., for each $t$ if $r_i \in \dom(lst(t))$ we have that
$({\it lst}(t))(r_i) = l_i$ provided $r_i := l_i$ appears in $\Init$.
We let ${\it lst}_\Init$ be the local state compatible with $\Init$
and let $\Gamma_\Init = ({\it lst}_\Init,\gamma_\Init, \beta_\Init)$.


\begin{figure*}[t]
  \centering \small
  $\inference[{\sc Read}] {a \in \{rd(x, n), rd^\mathsf{A}(x, n) \}
    \qquad (w, q) \in \gamma.\OW(t, x) \qquad
    \wrval(w) = n \\
    \tview' = \ensuremath{
      \begin{cases}
        \gamma.\tview_t\otimes\gamma.\mview_{(w,q)}&\mbox{if
          $(w, a) \in \WR \times \RA$ 
         }\\
         \gamma.\tview_t[x := (w, q)]&\mbox{otherwise}
       \end{cases}} 
     \\    \ctview' = \ensuremath{
      \begin{cases}
        \beta.\tview_t\otimes\gamma.\mview_{(w,q)}&\mbox{if
          $(w, a) \in \WR \times \RA$ 
         }\\
         \beta.\tview_t &\mbox{otherwise}
      \end{cases}} 
    }
    {\gamma, \beta\  \strans{a}_{t}\  \gamma[\tview_t \asgn \tview'], \beta[\tview_t \asgn \ctview'] }$
    \bigskip
    
  $ \inference[{\sc Write}] {
    a \in \{ wr(x,n), wr^{\sf R}(x,n)\} \qquad (w, q) \in \gamma.\OW(t, x) \setminus \gamma.\covered \\  \fresh_\gamma(q,q') \qquad
    \writes' = \gamma.\writes \cup \{(a, q')\} \\ 
    \tview' = \gamma.\tview_t[x := (a, q')] \qquad \mview' = \tview' \cup \beta.\tview_t
    }
    {\gamma, \beta\  \strans{a}_{t}\  \gamma[\tview_t \asgn \tview', \mview_{(a,q')} \asgn \mview', \writes \asgn \writes'], \beta }$
    \bigskip
    
  $
    \inference[{\sc Update}] {
    a = upd^{\sf
      RA}(x,m,n) 
    \qquad (w, q) \in \gamma.\OW(t, x) \setminus \gamma.\covered
    \qquad
    \wrval(w) = m 
    \\
     \fresh_\gamma(q,q') \qquad \writes' = \gamma.\writes \cup \{(a, q')\} \\
    \covered' = \gamma.\covered \cup \{(w, q)\} \qquad \mview' = \tview' \cup \ctview'
    \\
    \tview' = \ensuremath{
      \begin{cases}
        \gamma.\tview_t[x \asgn (a,
        q')]\otimes\gamma.\mview_{(w,q)}&\mbox{if $w \in
          \WR$ 
        }\\
        \gamma.\tview_t[x \asgn (a, q')]&\mbox{otherwise}
      \end{cases}}
    \\ 
    \ctview' = \ensuremath{
      \begin{cases}
        \beta.\tview_t \otimes\gamma.\mview_{(w,q)}&\mbox{if $w \in
          \WR$ 
        }\\
        \beta.\tview_t &\mbox{otherwise}
      \end{cases}}     
  }
    {\gamma, \beta\ \strans{a}_{t}\  \gamma\left[
      \begin{array}[c]{@{}l@{}}
        \tview_t \asgn \tview', \mview_{(a,q')} \asgn \mview', \\
        \writes \asgn \writes',
        \covered \asgn \covered'
      \end{array}\right], \beta[\tview_t\asgn \ctview']}$
\caption{Transition relation for reads, writes and updates of the
  memory semantics}
  \label{fig:surrey-opsem}
\end{figure*}

\smallskip
\noindent{\bf Transition semantics.}
The transition relation of our semantics for global reads and writes
is given in \reffig{fig:surrey-opsem} and builds on an earlier
semantics that does not distinguish the state of the
context~\cite{ECOOP20}. Each transition
$\gamma, \beta \strans{a}_{t} \gamma', \beta'$ is labelled by an action
$a$ and thread $t$ and updates the target state $\gamma$ (the state of
component being executed) and the context
$\beta$. 

\smallskip
\noindent{\bf {\sc Read} transition by thread $t$.} Assume that $a$ is
either a relaxed or acquiring read to variable $x$, $w$ is a write to
$x$ that $t$ can observe (i.e., $(w, q) \in \gamma.\OW(t,x)$), and the
value read by $a$ is the value written by
$w$. 
Each read causes the viewfront of $t$ to be updated. For an
unsynchronised read, $\tview_t$ is simply updated to include the new
write. A synchronised read causes the executing thread's view of the
executing component and context to be updated. In particular, for each
variable $x$, the new view of $x$ will be the later (in timestamp
order) of either $\tview_t(x)$ or $\mview_w(x)$. To express this, we
use an operation that combines two views $V_1$ and $V_2$, by
constructing a new view from $V_1$ by taking the later view of each
variable: \medskip

\hfill $  \begin{array}[t]{@{}l@{}l}
V_1 \otimes V_2  = \lambda x \in \dom(V_1).\ & 
    \textbf{if}~\tst(V_2(x)) \leq \tst(V_1(x))\  \textbf{then}\ V_1(x)\ \textbf{else}\  V_2(x)
  \end{array}
$\hfill {}


\smallskip
\noindent{\bf {\sc Write} transition by thread $t$.} A write
transition must identify the write $(w,q)$ after which $a$
occurs. This $w$ must be observable and must {\em not} be covered ---
the second condition preserves the atomicity of read-modify-write
updates. We must choose a fresh timestamp $q' \in \rat$ for $a$, which
for a C11 state $\gamma$ is formalised by $\fresh_\gamma(q, q') = q < q' \wedge \forall w' \in \gamma.\writes.\ q < \tst(w') \Rightarrow q' < \tst(w')$. 
That is, $q'$ is a new timestamp for variable $x$ and that
$(a,q')$ occurs immediately after $(w,q)$. The new write is added to
the set $\writes$. 

We update $\gamma.\tview_t$ to include the new
write, which ensures that $t$ can no longer observe any writes prior to
$(a, q')$. Moreover, we set the viewfront of $(a, q')$ to be the new
viewfront of $t$ in $\gamma$ together with the thread viewfront of the
environment state $\beta$. If some other thread synchronises with this
new write in some later transition, that thread's view will become at
least as recent as $t$'s view at this transition. Since $\mview$ keeps
track of the executing thread's view of both the component being
executed and its context, any synchronisation through this new write
will update views across components.

\smallskip
\noindent{\bf {\sc Update} transition by thread $t$.} These
transitions are best understood as a combination of the read and write
transitions. As with a write transition, we must choose a valid fresh
timestamp $q'$, and the state component $\writes$ is updated in the
same way. State component $\mview$ includes information from the new
view of the executing thread $t$.  As discussed earlier, in {\sc
  Update} transitions it is necessary to record that the write that
the update interacts with is now covered, which is achieved by adding
that write to $\covered$. Finally, we must compute a new thread view,
which is similar to a {\sc Read} transition, except that the thread's
new view always includes the new write introduced by the
update. 


\section{Abstract object semantics}
\label{sec:abstr-object-semant}
The rules in \reffig{fig:surrey-opsem} provide a semantics for read,
write and update operations for component programs within an executing
context and can be used to model clients and libraries under RC11
RAR. These rules do not cover the behaviour of abstract objects, which
we now consider.


There have been many different proposals for specifying and verifying
concurrent objects
memory~\cite{DBLP:conf/pldi/Kokologiannakis19,DBLP:conf/popl/BattyDG13,DBLP:journals/pacmpl/EmmiE19,DBLP:conf/esop/KrishnaEEJ20,DBLP:journals/pacmpl/RaadDRLV19,ifm18,DongolJRA18},
since there are several different objectives that must be addressed.
These objectives are delicately balanced in
linearizability~\cite{HeWi90}, the most well-used consistency
condition for concurrent objects. Namely, linearizability ensures:
\begin{enumerate*}
\item The abstract specification is explainable with respect to a
  \emph{sequential} specification.
\item Correctness is \emph{compositional}, i.e., any concrete
  execution of a system  comprising two linearizable objects is
  itself linearizable.
\item Correctness ensures \emph{contextual (aka observational)
    refinement}, i.e., the use of a linearizable implementation within
  a client program in place of its abstract specification does not
  induce any new behaviours in the client program.
\end{enumerate*}

There is however an inherent cost to linearizability stemming from the
fact that the effect of each method call must take place \emph{before}
the method call returns. In the context of weak memory, this
restriction induces additional synchronisation that may not
necessarily be required for
correctness~\cite{DBLP:journals/cacm/SewellSONM10}. Therefore, over
the years, several types of relaxations to the above requirements have
been proposed
~\cite{DBLP:conf/pldi/Kokologiannakis19,DBLP:conf/popl/BattyDG13,DBLP:journals/pacmpl/EmmiE19,DBLP:conf/esop/KrishnaEEJ20,DBLP:journals/pacmpl/RaadDRLV19,ifm18,DongolJRA18}.




General data structures present many different design choices at the
abstract level~\cite{DBLP:journals/pacmpl/RaadDRLV19}, but discussing these now detracts from our main
contribution, i.e., the integration and verification of clients and
libraries in a weak memory model. 
Therefore, we restrict our attention to an abstract lock object, which
is sufficient to highlight the main ideas. Locks have a clear ordering
semantics (each new lock $\acquire$ and lock $\release$ operation must
have a larger timestamp than all other existing operations) and
synchronisation requirements (there must be a release-acquire
synchronisation from the lock \emph{\release} to the lock
\emph{\acquire}).

To enable proofs of contextual refinement (see
\refsec{sec:cont-refin}), we must ensure corresponding method calls
return the same value at the abstract and concrete levels. To this
end, we introduce a special variable $rval$ to each local state that
stores the value that each method call returns.
\begin{example}[Abstract lock]
  \label{ex:abs-lock}
  Consider the specification of a lock with methods 
  {\tt Acquire}, and {\tt Release}. 
  Each method call of the lock is indexed by a subscript to uniquely
  identify the method call.  For the lock, the subscript is a counter
  indicating how many lock operations have been executed and is used
  in the example proof in \refsec{sec:example-verification}.
  \[ 
    \inference[{\sc Acquire}] {a = l.\acquire_n \quad ls' = ls[rval := \True] 
    } 
    {({\tt l.Acquire()}, ls) \trans a (\True, ls')}
  \]
  \[ \inference[{\sc Release}]
    {a = l.\release_n \quad ls' = ls[rval := \bot]} 
    {({\tt l.Release()}, ls) \trans a (\bot, ls')}
  \]

  Locks, by default are synchronising. That is, in the memory
  semantics, a (successful) \emph{\acquire} requires the operation to
  synchronise with most recent lock \emph{\release} (in a manner
  consistent with release-acquire semantics), so that any writes that
  are happens-before ordered before the \emph{\release} are visible to
  the thread that acquires the lock. The initial state of an abstract
  lock $l$, $\beta_\Init$, is given by: 
  \begin{align*}
    \beta_\Init.\writes & = \{(l.init_0,0)\}
    & 
    \gamma_\Init.\tview_t(l) & =  (l.init_0,0)
    \\
    \gamma_\Init.\covered & =  \emptyset
  \end{align*}  
  We also obtain the rules below, where we assume $\gamma$ is the state of
  the lock and $\beta$ is the state of the client.
  \begin{figure*}[t]
    \centering \small
    $    \inference[{\sc Acquire}] {
      a = l.\acquire_n  \qquad b = l.\acquire_n(t) \qquad (w, q) \in \gamma.\writes \qquad  w \in \{l.init_0, l.\release_{n-1}\} \\
      q = \maxTS(l,\gamma) \quad q < q' \\ \writes' = \gamma.\writes
      \cup \{(b, q')\} 
      \qquad \mview' = \tview' \cup \ctview' \qquad \covered' = \sigma.\covered \cup \{(w,q)\} \\
      \tview' = \gamma.\tview_t[l \asgn (b,
      q')]\otimes\gamma.\mview_{(w,q)} \quad \ctview' = \beta.\tview_t
      \otimes\gamma.\mview_{(w,q)} \\
      } {\gamma, \beta\ \strans{a}_{t}\
      \gamma\left[
        \begin{array}[c]{@{}l@{}}
          \writes := \writes', \tview_t := \tview', 
          \\
          \mview_{(b, q')} := \mview', \covered := \covered'
        \end{array}\right]
      , \beta[\tview_t \asgn \ctview']}
    $

    \bigskip
    $
    \inference[{\sc Release}] {
        a = l.\release_n \qquad w = l.\acquire_{n-1}(t) \qquad 
        (w, q) \in \gamma.\writes \qquad
        q = \maxTS(l,\gamma) \qquad q < q'  \\ 
        \writes' = \gamma.\writes \cup \{(a, q')\} \qquad
        \tview' = \gamma.\tview_t[x \asgn (a, q')]
        \qquad 
        \mview' = \tview' \cup \beta.\tview_t
      } 
      {\gamma, \beta\  \strans{a}_{t}\  \gamma\left[
          \begin{array}[c]{@{}l@{}}
            \writes := \writes', \tview_t := \tview', 
            \mview_{(a, q')} := \mview'
          \end{array}\right]
        , \beta}
      $
    \caption{Operational semantics for lock acquire and release }
    \label{fig:acq-rel-sem}
  \end{figure*}

  To record the thread that currently owns the lock, we derive a new
  action, $b$, from the action $a$ of the program
  semantics. Action $(w, q)$ represents the method that is observed by
  the $\acquire$ method, which must be an operation in
  $\gamma.\writes$ such that $q$ has the maximum timestamp for $l$
  (i.e., $q = \maxTS(l,\gamma)$. The new timestamp $q'$ must be larger
  than $q$. We create a new component state $\gamma'$ from $\gamma$ by
  \begin{itemize}[leftmargin=*]
  \item inserting $(b, q')$ into $\gamma.\writes$;
  \item updating $\tview_t$ to $\tview'$, where $\tview'$
    synchronises with the previous thread view in $\gamma$ to include
    information from the modification view of $(w, q)$, and updates $t$'s view of $l$ to include the new operation $(b, q')$; 
  \item updating the contextual thread view for $t$ to $\ctview'$,
    where $\ctview'$ synchronises with the previous thread view in the
    context state $\beta$ to include information from the modification
    view of $(w, q)$; and
  \item updating the modification view for the new operation $(b, q')$
    to $\mview'$, where $\mview'$ contains the view of $t$.
  \end{itemize}
  Finally, the context state $\beta'$ updates the thread view of $t$ to
  $\ctview'$ since synchronisation with a release may cause the view
  to be updated.
  
  A lock release, simply introduces a new operation with a maximal
  timestamp, provided that the thread executing the release currently holds
  the lock.
\end{example}






\newcommand{\lockver}{{\it lv}}
\section{Client-library verification}
\label{sec:example-verification}

Having formalised the semantics of clients and libraries in a weak
memory setting, we now work towards verification of
(client) programs that use such libraries.



\subsection{Assertion language}
\label{sec:assertion-language}
In our proof, we use \emph{observability assertions}, which describe
conditions for a thread to observe a specific value for a given 
variable. Unlike earlier works, our operational semantics covers
clients and their libraries, and hence operates over pairs of
states. 

 \smallskip
\noindent
{\bf Possible observation}, denoted $\langle x = u\rangle_t$, means
that thread $t$ \emph{may} observe value $u$ for
$x$~\cite{ECOOP20}. We extend this concept to cope with abstract
method calls as follows. In particular, for an object $o$ and method
$m$, we use $\langle o.m \rangle_t$ to denote that thread $t$ can observe
$o.m$.
  \begin{align*}
    \langle x = n \rangle_t (\sigma)  & \ \  \equiv \ \ \exists w\in
                              \sigma.\OW(t,x).\ \wrval(w) = n\\
    \langle o.m \rangle_t (\sigma) & \ \  \equiv \ \ \exists q.\ (o.m, q) \in
                            \sigma.\writes \wedge q \geq \sigma.\tview_t(o)
  \end{align*}
  
  \noindent To distinguish possible observation in clients and libraries, we introduce the following notation,
  where $\gamma$ and $\beta$ are the client and library states,
  respectively, and $p$ is either a valuation (i.e., $x = n$) or an
  abstract method call (i.e., $o.m$):
  \begin{align*}
    \langle p \rangle_t^C (\gamma, \beta) & \ \ \equiv \ \ \langle p \rangle_t (\gamma)  &
                                                                                               \langle p \rangle_t^L (\gamma, \beta) & \ \ \equiv \ \ \langle p \rangle_t(\beta)
  \end{align*}
  
\smallskip
\noindent
  {\bf Definite observation}, denoted $[x = u]_t$, means that
  thread $t$ can only see the last write to $x$, and that this write
  has written value $u$. We define the \emph{last
    write} to $x$ in a set of writes $W$ as:
  \begin{align*}
    \last(W, x) = w \ \  \equiv \ \  &
                                       \begin{array}[t]{@{}l@{}}
                                         w \in \{w \in W \mid \var(w) = x \} \wedge {}
                                         \\
                                         (\forall w' \in W_{|x}.\ \tst(w') \leq \tst(w))
                                       \end{array}
  \end{align*}
  We define the definite observation of a view function, $view$ with
  respect to a set of writes as follows:
  \begin{align*}
    \begin{array}[t]{@{}r@{}l@{}}
      & \dview (view, W, x) = n \\
      \ \   \equiv\ \  &  view(x) = \last(W, x) \wedge 
    \wrval(\last(W, x)) = n
    \end{array}
  \end{align*}
  The first conjunct ensures that the viewfront of $view$ for $x$ is the
  last write to $x$ in $W$, and the second conjunct ensures that the
  value written by the last write to $x$ in $W$ is $n$.  For a variable
  $x$, thread $t$ and value $n$, we define:
  \begin{align*}
    [x = n]_t (\sigma) \ \ & \equiv \ \ \dview(\sigma.\tview_t, \sigma.\writes \cap \W, x) =n
  \end{align*}
  The extension of definite observation assertions to abstract method
  calls is straightforward to define. Namely we have: 
  \begin{align*}
    [o.m]_t (\sigma) \ \ & \equiv \ \ 
                  \begin{array}[t]{@{}l@{}}
                    \sigma.\tview_t(o) =  \maxTS(o, \sigma) \wedge {} \\
                    (o.m, \maxTS(o, \sigma)) \in \sigma.\writes
                  \end{array}
  \end{align*}
  As with possible observations, we lift definite observation
  predicates to state spaces featuring clients and libraries: 
  \begin{align*}
    [p]^C_t (\gamma, \beta) & \ \ \equiv \ \ [p]_t(\gamma) & 
                                                                 [p]^L_t (\gamma, \beta) & \ \ \equiv \ \ [p]_t(\beta)
  \end{align*}

\smallskip
\noindent {\bf Conditional observation}, denoted
$\CObs{x}{u}{t}{y}{v}$, means that if thread $t$ synchronises with a
write to variable $x$ with value $u$, it \emph{must} subsequently
observe value $v$ for $y$.  For variables $x$ and $y$, thread $t$ and
values $u$ and $v$, we define:
\begin{align*}
                         \begin{array}[t]{@{}r@{~}l@{}}
                           & \CObs{x}{u}{t}{y}{v} (\sigma) \\
                           \ \ \equiv \ \ & 
                           \forall w \in \sigma.\OW(t,x) .\
                           \wrval(w) = u \imp  \\ 
                           & \quad \act(w) \in \WR \wedge \dview(\sigma.\mview_w, \sigma.\writes, y) = v 
                         \end{array}
\end{align*}
This is a key
assertion used in message passing proofs
\cite{ECOOP20,DBLP:conf/ecoop/KaiserDDLV17} since it guarantees an
observation property on a variable, $y$, via a synchronising read of
another variable, $x$.


As with possible and definite assertions, conditional assertions can
generalised to objects and extended to pairs of states describing a
client and its library. However, unlike possible and definition
observations assertions, conditional observation enables one to
describe view synchronisation across different states. For example,
consider the following, which enables conditional observation of an
abstract method to establish a definite observation assertion for the
thread view of the client.  We assume a set $Sync \subseteq \Act$ that
identifies a set of synchronising abstract actions.
\begin{align*}
  \begin{array}[t]{@{}r@{~}l@{}}
    & \langle o.m \rangle^L[y = v]_t^C (\gamma, \beta)\\  \ \ \equiv &
                                                                          o.m \in Sync \wedge
    \forall q.\ (o.m, q) \in \sigma.\writes \wedge q \geq \gamma.\tview_t(o) \imp \\
&       {} \qquad\qquad  \dview(\gamma.\mview_{(o.m, q)}, \beta.\writes, y) = v
    \end{array}
\end{align*}
It is possible to define other variations, e.g., conditional observation
synchronisation from clients to libraries, but we leave out the
details of these since they are straightforward to construct. 

\smallskip
\noindent {\bf Covered operations}, denoted $\cvd{x}{u}$, where $x$ is
a variable and $u$ a value.  Recall from the {\sc Acquire} rule that
a new acquire operation causes the immediately prior (release)
operation $l.release_{n-1}$ to be covered so that no later acquire can
be inserted between $l.release_{n-1}$ and the new acquire. 
To reason about this phenomenon over states, we use:
\begin{align*}
  \cvd{o.m}{} (\sigma) &\ \  \equiv\ \
                         \begin{array}[t]{@{}l@{}}
                           \forall (w, q) \in \sigma.\writes_{|o} \setminus \sigma.\covered.\  \\
                           w = o.m \wedge q = \maxTS(o, \sigma)
                         \end{array}
\end{align*}
where $\sigma.\writes_{|o}$ is the set of operations over object $o$.

\smallskip
\noindent {\bf Hidden value}, denoted $\cvv{o.m}{}$, states that
the operation $o.m$ exists, but all of these are hidden from
interaction. 
In proofs, such assertions limit the values that
can be returned. 
\begin{align*}
  \cvv{o.m}{}(\sigma)
  &\ \  \equiv\ \  
    \begin{array}[t]{@{}l@{}}
      (\exists q.\ (o.m, q) \in \sigma.\writes) \wedge {}\\
      (\forall q.\ (o.m, q) \in \sigma.\writes  \imp (o.m, q) \in \sigma.\covered)
    \end{array} 
\end{align*}
Both covered and hidden-value assertions can be lifted to pairs of
states and can be used to reason about standard writes, as opposed to
method calls (details omitted).

\subsection{Hoare Logic for C11 and Abstract Objects}
\label{sec:hoare-logic-c11}
Since we have an operational semantics, the assertions in \refsec{sec:assertion-language} can be
integrated into standard Hoare-style proof calculus in a
straightforward
manner~\cite{ECOOP20,DBLP:journals/corr/abs-2004-02983}. The only
differences are the state model (which is a weak memory state, as
opposed to mappings from variables to values) and the atomic
components (which may include reads of global variables, and, in this
paper, abstract method calls).

Following \cite{ECOOP20,DBLP:journals/corr/abs-2004-02983}, we let
$\Sigma_C$ and $\Sigma_{L}$ to be the set of all possible global
state configurations of the client and library, respectively and let
$\Sigma_{C11} = (\LVar \to \Val) \times \Sigma_C \times \Sigma_{L}$
be the set of all possible client-library C11 states. Predicates over
$\Sigma_{C11}$ are therefore of
type $\Sigma_{C11} \to \mathbb{B}$.  This leads to the following
definition of a Hoare triple, which we note is the same as the
standard definition --- the only difference is that the state
component is of type $\Sigma_{C11}$.
\begin{definition}
  \label{def:soundn-class-rules-1}Suppose 
  $p, q \in \Sigma_{C11} \to \mathbb{B}$, $P \in {\it Prog}$ and
  $\textsf{\textbf{E}} = \lambda t : \Tid.\ \kwskip$.  The semantics
  of a Hoare triple under partial correctness is given by:
  
  $\begin{array}{r@{~}l}
    \{p\} \Init \{q\} & =  q(\Gamma_\Init) 
    \\[2pt]
        \{p\}
\Init ; P \{q\} & =  \exists r.\,\{p\} 
                      \Init \{r\} \wedge \{r\} P \{q\}

     \\[2pt]
    \multicolumn{2}{l}{
                   \{p\}P \{q\} = \begin{array}[t]{@{}l@{}}
                     \forall \lstate,\gamma, \beta ,\lstate',\gamma', \beta'.\  p(\lstate,
                   \gamma, \beta) \wedge \\
                     \   (P, \lstate, \gamma, \beta) \ltsArrow{}^* (\textsf{\textbf{E}},
                   \lstate', \gamma', \beta') \imp q(\lstate',\gamma', \beta')
                   \end{array}}
    \\
  \end{array}$
\end{definition}
This definition (in the context of RC11
\cite{DBLP:conf/pldi/LahavVKHD17}) allows all standard Hoare logic
rules for compound statements to be reused \cite{ECOOP20}.  Due to
concurrency, following Owicki and Gries, one must prove \emph{local
  correctness} and \emph{interference freedom} (or
stability)~\cite{DBLP:journals/acta/OwickiG76,ECOOP20,DBLP:journals/corr/abs-2004-02983,DBLP:conf/icalp/LahavV15}. This
is also defined in the standard manner.
Namely, a statement $R \in \AComm$ with precondition $pre(R)$ (in the
standard proof outline) does {\em not interfere} with an assertion $p$
iff $\{ p \wedge pre(R)\} \ R \ \{ p\}.$ Proof outlines of concurrent
programs are {\em interference free} if no statement in one thread
interferes with an assertion in another thread.

The only additional properties that one must define are on the
interaction between atomic commands and predicates over assertions
defined in \refsec{sec:assertion-language}. A collection of rules for
reads, writes and updates have been given in prior
work~\cite{DBLP:journals/corr/abs-2004-02983,ECOOP20}. Here, we
present rules for method calls of the abstract lock object defined in
\refex{ex:abs-lock}.

In proofs, it is often necessary to reason about particular versions
of the lock (i.e., the lock counter). Therefore, we use ${\tt l.Acquire(\mbox{$v$})}$
and ${\tt l.Release(\mbox{$v$})}$ to denote the transitions that set the lock version to
$v$. Also note that in our example proof, it is clear from context
whether an assertion refers to the client or library state, and hence,
for clarity, we drop the superscripts $C$ and $L$ as used in
\refsec{sec:assertion-language}.

The lemma below has been verified in Isabelle/HOL.

\begin{lemma} Each of the following holds,  
  where the statements are
  decorated with the  identifier of the executing thread, assuming ${\tt m} \in \{{\tt Acquire}, {\tt Release}\}$ and $t \neq t'$
  \begin{small}
  \begin{align} 
  & \{\cvv{l.\release_u}{}\}\ {\tt
    l.Acquire(\mbox{$v$})_t} \ \{ v > u + 1 \} 
    \\
   & \{\cvv{l.\release_u}{}\}\ {\tt l.m(\mbox{$v$})_t}\
    \{\cvv{l.\release_u}{}\} 
    \\
   & \{[l.\release_u]_t\}
    \ {\tt
    l.Acquire(\mbox{$v$})_t}\ \{[l.\acquire_{u+1}]_t\}\\
   & \{[x = u]_t\}\ {\tt l.m(\mbox{$v$})_{t'}}\
    \{[x = u]_t\} \\
   & \{
       \langle l.\release_u \rangle [x = n]_t \}\
     {\tt l.Acquire(\mbox{$v$})_t}\ \{v = u + 1 \imp [x = n]_t\} 
    \\
   &  \left\{
     \begin{array}[c]{@{}l@{}}
       \neg \langle l.\release_u \rangle_{t'} 
        {}\wedge{} {[}x = v]_t
     \end{array}
\right\}\ {\tt l.Release(\mbox{$u$})_{t}}\
     \{\langle \release_u \rangle [x = v]_{t'}\} 
     \label{co-intro}
  \end{align}
\end{small}


\end{lemma}
  








\subsection{Example Client-Library Verification}
\label{sec:example-client-lbjec}
To demonstrate use of our logic in verification, consider the simple
program in \reffig{fig:lockmp-proof}, which comprises a lock object
$l$ and shared client variables $d_1$ and $d_2$ (both initially~$0$).
Thread~1 writes 5 to both $d_1$ and $d_2$ after acquiring the lock
while thread~2 reads $d_1$ and $d_2$ (into local registers $r_1$ and
$r_2$) also after acquiring the lock.

Under SC, it is a standard exercise to show that
the program terminates with $r_1 = r_2$ and $r_i = 0$ or $r_i =
5$. 
We show that the lock specification in
\refsec{sec:abstr-object-semant} together with the assertion language
from \refsec{sec:assertion-language} and Owicki-Gries logic from
\refsec{sec:hoare-logic-c11} is sufficient to prove this property. In
particular, the specification guarantees {\em adequate synchronisation} so
that if the Thread~2's lock acquire sees the lock release in Thread 1,
it also sees the writes to $d_1$ and $d_2$. The proof 
relies on two distinct types of properties:
\begin{itemize}[leftmargin=*]
\item \emph{Mutual exclusion}: As in SC, no
  two threads should execute their critical sections at the same
  time.
\item \emph{Write visibility}: If thread 1 enters its critical section
  first, its writes to both $d_1$ and $d_2$ must be visible to thread
  2 after thread 2 acquires the lock. Note that this property is not
  necessarily guaranteed in a weak memory setting since all accesses
  to $d_1$ and $d_2$ in \reffig{fig:lockmp-proof} are relaxed.
\end{itemize}

\begin{figure*}[t]
  \centering
\fbox{
\begin{minipage}[b]{0.7\textwidth}
    \begin{center}  \small
  {\bf Init: } $d_1:=0;$ $d_2:=0;$ ${\tt l.init()};$  \qquad \qquad \qquad \qquad\\ 
   {\color{green!40!black} $\{Inv \wedge [d_1 = 0]_1 \wedge [d_2 = 0]_1 \wedge [d_1 = 0]_2 \wedge [d_2 = 0]_2\}$} \\ 
  $\begin{array}{l@{\quad}||@{\quad }l}
     \text{\bf Thread } 1
     & \text{\bf Thread } 2\\
     
     \begin{array}[t]{@{}l@{}}
       1: \{Inv \wedge \mathbf{P_1}\}\ {\bf if}\ {\tt l.Acquire()} \\ 
       2: \{Inv \wedge \mathbf{P_2}\}\ \quad d_1 := 5 ; \\
       3: \{Inv \wedge \mathbf{P_3}\}\  \quad d_2 := 5 ; \\
       4: \{Inv \wedge \mathbf{P_4}\}\ \quad {\tt l.Release()} 
     \end{array}
     & 
     \begin{array}[t]{@{}l@{}}
       1: \{Inv \wedge \mathbf{Q_1}\}\ {\bf if}\ {\tt l.Acquire(\mbox{$rl$})} \\ 
       2: \{Inv \wedge \mathbf{Q_2}\}\ \quad r_1 \gets d_1 ; \\
       3: \{Inv \wedge \mathbf{Q_3}\}\ \quad r_2 \gets d_2 ; \\
       4: \{Inv \wedge \mathbf{Q_4}\}\ \quad{\tt l.Release()} 
     \end{array}
     \end{array}$

   {\color{green!40!black} $5: \{(r_1=0 \wedge r_2=0) \vee (r_1=5 \wedge r_2=5)\}$}  \qquad \qquad  \qquad      
 \end{center}
\end{minipage}}
\medskip

\raggedright \small
where assuming $P_{po}  =  (pc_2 = 1 \imp  \neg \langle l.\release_2 \rangle_2)\wedge \cvv{l.\init_0}{}$, we have  

\centering
$
\begin{array}[t]{@{}l@{\qquad}l@{}}
 \begin{array}[t]{@{}r@{}l@{}}
   P_1 =  &   [d_1 = 0]_1
     \wedge [d_2 = 0]_1 
           \wedge  \\
  & (pc_2 = 1 \imp [l.init_0]_{1} \wedge [l.init_0]_{2})\\
  &  \wedge  (pc_2 \in \{2,3,4\} \imp \cvd{l.\acquire_1}{}) \\
  P_2 =   & [d_1 = 0]_1 \wedge [d_2 = 0]_1
   \wedge P_{po}
\\
  P_3 =  & [d_1 = 5]_1 \wedge [d_2 = 0]_1
   \wedge P_{po}
        \\
  P_4  =  & [d_1 = 5]_1 \wedge [d_2 = 5]_1
   \wedge
   P_{po}
   \\
   \end{array}
        & 
          \begin{array}[t]{@{}r@{}l@{}}
            & Q_1' = pc_1 = 5 \wedge \langle l.\release_2 \rangle [d_1 = 5]_2 {} \wedge \langle l.\release_2\rangle [d_2 = 5]_2 \\
     & Q_1  =  \begin{array}[t]{@{}l@{}}
   \left(
   \begin{array}[c]{@{}l@{}l@{}}
     pc_1 \notin \{2,3,4\} \imp & ([l.init_0]_2 \wedge [d_1 = 0]_2 \wedge [d_2 = 0]_2) \lor Q_1' {} \\
   \end{array}\right)
   \\
   {} \wedge (pc_1 = 1 \imp [l.init_0]_1) 
   \wedge (pc_1 = 5 \imp \cvv{l.\init_0}{})
 \end{array} 
\\
 & Q_2  =  (rl = 1 \imp [d_1 = 0]_2 \wedge [d_2 = 0]_2)
    \\
 & \qquad\ \  {} \wedge (rl = 3 \imp [d_1 = 5]_2 \wedge [d_2 = 5]_2)   
\\
 & Q_3  = (rl = 1 \imp r_1 = 0 \wedge [d_2 = 0]_2) \\
& \qquad\ \ {}   \wedge (rl = 3 \imp r_1 = 5 \wedge [d_2 = 5]_2)   
        \\
 & Q_4  = (rl = 1 \imp r_1 = 0 \wedge r_2 = 0)
        \wedge (rl = 3 \imp r_1 = 5 \wedge r_2 = 5)
        \end{array}
\end{array}
$

 \caption{Proof outline for lock-synchronisation}
\label{fig:lockmp-proof}
\end{figure*}

Our proof is supported by the following global invariant:
\begin{align*}
  Inv \ \ \equiv{}\ \  & \neg (pc_1 \in \{2,3,4\} \wedge pc_2 \in \{2,3,4\}) \wedge (rl \in \{1, 3\})
\end{align*}
The first conjunct establishes mutual exclusion, while the second
ensures that the lock version written by the acquire in thread 2 is
either 1 or 3, depending on which thread enters its critical section
first.



The main purpose of the
definite and possible observation assertions is to establish 
$Q_1'$ (which appers in $Q_1$) using rule \refeq{co-intro}.
This predicate helps establish $[d_1 = 5]_2$ and
$[d_2 = 5]_2$ in thread 2 whenever thread 2 acquires the lock after
thread~1. 
    
    
    
    
The most critical of these assertions is $Q_1$, which states that if
thread 1 is not executing it's critical section then we either have
\begin{itemize}[leftmargin=*]
\item $[l.init_0]_2 \wedge [d_1 = 0]_2 \wedge [d_2 = 0]_2$, i.e.,
  thread 2 can definitely see the lock initialisation and definitely
  observes both $d_1$ and $d_2$ to have value $0$, or
\item 
  $Q_1'$ holds,
  i.e., thread 1 has
  released the lock and has established a state whereby if thread 2
  acquires the lock, it will be able to establish the definite value
  assertions $[d_1 = 5]_2$ and $[d_2 = 5]_2$.
\end{itemize}
Note that $Q_1$ also includes a conjunct
$pc_1 = 5 \imp \cvv{l.\init_0}{}$, which ensures that if thread 2 enters
its critical section after thread 1 has terminated, then it does so
because it sees $l.\release_2$ (as opposed to $l.\init_0$). This means
that we can establish $rl = 1 \imp [d_1 = 0]_2 \wedge [d_2 = 0]_2$
(i.e., thread 2 has acquired the lock first) and
$rl = 3 \imp [d_1 = 5]_2 \wedge [d_2 = 5]_2$ (i.e., thread 2 has
acquired the lock second) in $Q_2$. Using these definite value
assertions, we can easily establish that the particular values that
are loaded into registers $r_1$ and $r_2$. The lemma has been verified
in Isabelle/HOL.

\begin{lemma}
  \label{thm:lockmp-proof}
  The proof outline in \reffig{fig:lockmp-proof} is valid. 
\end{lemma}

\newcommand{\Tr}{{\it Tr_{SF}}}
\newcommand{\ct}{{\it ct}}
\newcommand{\at}{{\it at}}
\newcommand{\refby}{\sqsubseteq}
\newcommand{\crefby}{\mathrel{\widehat{\sqsubseteq}}}
\newcommand{\glo}{{\it glo}}
\newcommand{\st}{{\it state}}

\section{Contextual Refinement}
\label{sec:cont-refin}

We now describe what it
means to \emph{implement} a specification so that any client
properties that was preserved by the specification is not invalidated by the
implementation. 
We define
and prove contextual refinement directly, i.e., without appealing to
external correctness conditions over libraries, c.f.
linearizability~\cite{HeWi90,ifm18,GotsmanY11,DongolJRA18,DBLP:journals/tcs/FilipovicORY10}. 


\subsection{Refinement and Simulation for Weak Memory}
\label{sec:refin-simul-weak}

Since we have an operational semantics with an interleaving semantics
over weak memory states, the development of our refinement theory
closely follows the standard approach under
SC~\cite{DBLP:books/cu/RoeverE1998}.

Suppose $P$ is a program with initialisation $\Init$. An
\emph{execution} of $P$ is defined by a possibly infinite sequence
$\Pi_0\, \Pi_1\, \Pi_2\,\dots$ such that
\begin{enumerate}[leftmargin=*]
\item each $\Pi_i$ is a 4-tuple $(P_i, ls_i, \gamma_i, \beta_i)$
  comprising a program, local state, global client state and global library state, and
\item
  $(ls_0, \gamma_0, \sigma_0) = (\ls_\Init, \gamma_\Init,
  \sigma_\Init)$, and
\item for each $i$, we have $\Pi_i \Longrightarrow \Pi_{i+1}$ as
  defined in \refsec{sec:program-semantics}.
\end{enumerate}
A \emph{client trace} corresponding to an execution
$\Pi_0\,\Pi_1\,\Pi_2\dots$ is a sequence $\ct \in \Sigma_C^*$
such that $\ct_i = (\pi_{2}(\Pi_i)_{|C}, \pi_{3}(\Pi_i))$, where
$\pi_n$ is a projection function that extracts the $n$th component of
a given tuple and $ls_{|C}$ restricts the given local state $ls$ to
the variables in $\LVar_C$.  Thus each $\ct_i$ is the global client
state component of $\Pi_i$.

After a projection, the concrete implementation may contain (finite or
infinite) stuttering~\cite{DBLP:books/cu/RoeverE1998}, i.e.,
consecutive states in which the client state is unchanged. We let
${\it rem\_stut}(\ct)$ be the function that removes all stuttering
from the trace $\ct$, i.e., each consecutively repeating state is
replaced by a single instance of that state. We let $\Tr(P)$ denote
the set of \emph{stutter-free traces} of a program $P$, i.e., the
stutter-free traces generated from the set of all executions of $P$.

Below we refer to the client that uses the abstract object as the
\emph{abstract client} and the client that uses the object's
implementation as the \emph{concrete client}. The notion of contextual
refinement that we develop ensures that a client is not able to distinguish 
the use of a concrete implementation in place of an abstract
specification. In other words, each thread of the concrete client
should only be able to observe the writes (and updates) in the client
state (i.e., $\gamma$ component) that the thread could already observe
in a corresponding of the client state of the abstract client.

First we define trace refinement for weak memory states. 
\begin{definition}[State and Trace Refinement]
  \label{def:cont-refin-1}
  We say a concrete state $\gamma_C$ is a \emph{refinement} of an
  abstract state $\gamma_A$, denoted
  $(ls_A, \gamma_A) \sqsubseteq (ls_C, \gamma_C)$ iff $ls_A = ls_C$,
  $\gamma_A.\covered = \gamma_C.\covered$ and for all threads $t$ and
  $x \in \GVar$, we have
  $\gamma_C.\OW(t, x) \subseteq \gamma_A.\OW(t, x)$. 
  We say a concrete
  trace $\ct$ is a \emph{refinement} of an abstract trace $\at$,
  denoted $\at \refby \ct$, iff $\ct_i \sqsubseteq \at_i$ for all $i$.
\end{definition}
This now leads to a natural definition of contextual refinement that
is based on the refinement of traces.
\begin{definition}[Program Refinement]
  \label{def:prog-ref}
  A concrete program $P_C$ is a \emph{refinement} of an abstract
  program $P_A$, denoted $P_A \refby P_C$, iff for any (stutter-free)
  trace $\ct \in \Tr(P_C)$ there exists a (stutter-free) trace
  $\at \in \Tr(P_A)$ such that $\at \refby \ct$.
\end{definition}
Finally, we obtain a notion of contextual refinement for abstract
objects. Suppose $P$ is a program with holes. We let $P[O]$ be the
program in which the holes are filled with the operations from object
$O$. Note that $O$ may be an abstract object, in which case execution
of each method call follows the abstract object semantics
(\refsec{sec:abstr-object-semant}), or a concrete implementation, in
which case execution of each method call follows the semantics of
reads, writes and updates (\refsec{sec:program-semantics}).
\begin{definition}[Contextual refinement]
  \label{def:cref}
  We say a concrete object $CO$ is a \emph{contextual refinement} of
  an abstract object $AO$ 
   iff for any client
  program $C$, we have $C[AO] \refby C[CO]$.
\end{definition}

To verify contextual refinement, we use a notion of \emph{simulation},
which once again is a standard technique from the literature. The
difference in a weak memory setting is the fact that the refinement
rules must relate more complex configurations, i.e., tuples of the
form $(P, {\it lst}, \gamma, \alpha)$.

The simulation relation, $R$, relates triples
$(als, \gamma_A, \alpha)$, comprising an abstract local state $als$,
client state $\gamma_A$ and library state $\alpha$, with triples
$(cls, \gamma_C, \beta)$ comprising a concrete local state $cls$, a
client state $\gamma_C$ and concrete library state $\beta$. The
simulation condition must ultimately ensure
$(als_{|C}, \gamma_A) \refby (cls_{|C}, \gamma_C)$ at each step as
defined in \refdef{def:cont-refin-1}. However, since client
synchronisation can affect the library state, a generic forward
simulation rule is non-trivial to define since it requires one to
describe how clients steps affect the simulation relation. We
therefore present a simpler use case for libraries that are used by
clients that do not perform any synchronisation outside the library
itself (e.g., the client in \reffig{fig:lockmp-proof}). If
$\Pi = (P, lst, \gamma, \alpha)$, we let
$\st(\Pi) = (lst, \gamma, \alpha)$ be the state corresponding to
$\Pi$.


\begin{definition}[Forward simulation for synchronisation-free
  clients] \label{def:fsim} For an abstract object $AO$ and a concrete object $CO$ and
  a client $C$ that only synchronises through $AO$ (and $CO$),
  $C[AO] \refby C[CO]$ holds if there exists a relation $R$ such that
  \begin{enumerate}[leftmargin=*]
  \item $R((als, \gamma_A, \alpha), (cls, \gamma_C, \beta)) \imp $

    $
    \begin{array}[c]{l}
      als_{|C} = cls_{|C} \wedge \gamma_A.\covered = \gamma_C.\covered \wedge {} \\
      \forall t, x.\
      \begin{array}[t]{@{}l@{}}
\gamma_C.\OW(t, x) \subseteq \gamma_A.\OW(t, x) \wedge {} \\
      als(t)(rval) = cls(t)(rval)
      \end{array}
    \end{array}
$
  \hfill   (client observation)
  \item
    $R(\st(\Omega_\Init), \st(\Pi_\Init))$ \hfill (initialisation)
  \item For any concrete configurations $\Pi$, $\Pi'$ and abstract
    configuration $\Omega$, if $\Pi \Longrightarrow \Pi'$ via a step
    corresponding to $CO$, and $R(\st(\Omega), \st(\Pi))$, then either
    \begin{itemize}
    \item $R(\st(\Omega), \st(\Pi))$, or \hfill (stuttering step)
    \item there exists an abstract configuration $\Omega'$ such that
      $\Omega \Longrightarrow \Omega'$ and $R(\st(\Omega'),
      \st(\Pi'))$.
      
      \hfill (non-stuttering step)
    \end{itemize}
  \end{enumerate}
\end{definition}

\begin{theorem} If $R$ is a forward simulation between $AO$ and $CO$,
  then for any client that only synchronises through $AO$ (and $CO$)
  we have $C[AO] \refby C[CO]$.
\end{theorem}

\subsection{Sequence Lock}
\label{sec:sequence-lock}



The first refinement example is a sequence lock which operates over a single shared variable
($glb$). \smallskip

    \begin{minipage}[b]{0.80\textwidth}
      $\textbf{Init:} \ \ glb = 0$
      \\[6pt]
      \begin{minipage}[t]{\textwidth}
\begin{minipage}[b]{0.80\textwidth}
  \textbf{Acquire()}: 
        \begin{algorithmic}[1] \small
          \State \textbf{do} \quad \textbf{do}
            $r \leftarrow^{\sf A} glb$ 
          \textbf{until} $even (r)$ ;
          \State \qquad\ \   $loc \gets \kwcas(glb, r, r+1)$ 
          \State \textbf{until} $(loc)$
        \end{algorithmic}
      \end{minipage}
      \end{minipage}
      \\[6pt]     
      \begin{minipage}[t]{\textwidth}
        \begin{minipage}[b]{0.8\textwidth}
          \textbf{Release()}:
          \begin{algorithmic}[1] \small
            \State $glb :=^{\sf R} r + 2$ 
          \end{algorithmic}
        \end{minipage}
      \end{minipage}
    \end{minipage}\smallskip

\noindent
The ${\bf Acquire}$ operation returns true if, and only if, the
$\kwcas$ on line~2 is successful. Therefore, in order to prove the
refinement, we will need to prove that whenever the $\kwcas$ operation
is successful, the abstract object can also successfully acquire the
lock maintaining the simulation relation.  Also, the read on line 1
and the unsuccessful $\kwcas$ are stuttering steps and we need to show
that when those steps are taken the abstract state remains unchanged
and the new concrete state preserves the simulation relation. The
${\bf Release}$ operation contains only one releasing write on
variable $glb$, which is considered to be a refining step. It is
straightforward to show that this operation refines the abstract
object release operation. The  following proposition has been verified in
Isabelle/HOL.

\begin{proposition}
  For synchronisation-free clients, there exists a forward simulation
  between the abstract lock object and the sequence
  lock. 
\end{proposition}
\subsection{Ticket Lock}
\label{sec:ticket-lock}
Our second refinement example is the ticket lock:\medskip 

    \begin{minipage}[b]{0.48\textwidth}
      $\textbf{Init:} \ \ nt = 0, \ \ sn = 0$\\[6pt]
      \begin{minipage}[t]{\columnwidth}
 \begin{minipage}[b]{0.80\textwidth}
   \textbf{Acquire()}:
        \begin{algorithmic}[1] \small
          \State $m\_t \leftarrow \kwfai(nt)$
          \State  \textbf{do}
             $s\_n \leftarrow^{\sf A} sn$ 
             \textbf{until} $m\_t = s\_n$
        \end{algorithmic}
      \end{minipage}
      \end{minipage}
      \\[8pt]     
      \begin{minipage}[t]{\columnwidth}
        \begin{minipage}[b]{0.80\textwidth}
          \textbf{Release()}: 
        \begin{algorithmic}[1] \small
          \State $sn :v=^{\sf R} s\_n + 1$ 
        \end{algorithmic}
        \end{minipage}
      \end{minipage}
    \end{minipage}\medskip

    \noindent The ticket lock
has two shared variables $nt$ (next ticket) and $sn$ (serving
now). Invocation of {\bf Acquire} loads the next available ticket into
a local register ($m\_t$) and increases the value of $nt$ by one using
a fetch-and-increment ($\kwfai$) operation. It then enters a busy loop
and reads $sn$ until it sees its own ticket value in $sn$ before it
can enter its critical section.

If the read on line 2 of the {\bf Acquire} operation reads from a
write whose value is equal to the value of $m\_t$, then the
lock is acquired. Therefore we will need to show that if this
situation arises, the abstract lock object can also take a step and
successfully acquire the lock. We consider the $\kwfai$ operation on
line 1 and the read on line 2 if it reads a value that is not equal to
$m\_t$ to be a stuttering step. We prove that each of the
stuttering and non-stuttering steps preserves the simulation
relation. Similar to the previous example, the {\bf Release} operation
consists of only one releasing write to variable $sn$ and it is
straightforward to show that this operation refines the abstract
release operation. This proof has been mechanised in Isabelle/HOL. 

\begin{proposition}
  For synchronisation-free clients, there exists a forward simulation
  between the abstract lock object and the ticket
  lock. 
\end{proposition}



%

\section{Conclusions}
\label{sec:conclusion}

In this paper, we present a new approach to specifying and verifying
abstract objects over weak memory by extending an existing operational
semantics for RC11 RAR (which is a fragment of the C11 memory
model). We show that our methodology supports two types of
verification: (1) proofs of correctness of client programs that
\emph{use} abstract libraries and (2) refinement proofs between
abstract libraries and their implementations. Moreover, the
operational semantics allows one to execute programs in thread order
and accommodates weak memory behaviours via a special encoding of the
state. 
To exploit this operational
semantics, we develop an assertion language that describes a thread's
observations of client-library states, which is in turn used to verify
program invariants and proofs of refinement. The operational
semantics, proof rules and example verifications 
have been mechanised in Isabelle/HOL.


There are now several different approaches to program
verification that support different aspects of weak memory using
pen-and-paper proofs
(e.g.,~\cite{DBLP:conf/icalp/LahavV15,DBLP:conf/oopsla/TuronVD14,DBLP:conf/popl/AlglaveC17,doko2017tackling}),
model checking
(e.g.,~\cite{DBLP:conf/pldi/Kokologiannakis19,DBLP:conf/pldi/AbdullaAAK19}),
specialised tools
(e.g.,~\cite{DBLP:conf/pldi/TassarottiDV15,DBLP:conf/esop/KrishnaEEJ20,DBLP:conf/esop/SvendsenPDLV18,DBLP:conf/tacas/Summers018}),
and generalist theorem provers (e.g.,~\cite{ECOOP20}). These cover a
variety of (fragments of) memory models and proceed via exhaustive
checking, specialist separation logics, or Hoare-style calculi.

The idea that abstract methods should specify synchronisation
guarantees has been established in earlier
work~\cite{ifm18,DongolJRA18}, where it has been shown to be necessary
for contextual refinement~\cite{DongolJRA18} and
compositionality~\cite{ifm18}. Raad et al
\cite{DBLP:journals/pacmpl/RaadDRLV19} have tackled the problem of
client-library programs and also consider the C11 memory model.

Krishna et al \cite{DBLP:conf/esop/KrishnaEEJ20} have developed an
approach to verifying implementations of weakly consistent
libraries~\cite{DBLP:journals/pacmpl/EmmiE19}. They 
account for weak memory relaxations by transitioning over a generic
happens-before relation encoded within a transition system. On the one
hand, this means that their techniques apply to any memory model, but
on the other hand, such a happens-before relation must ultimately be
supplied. 

In future work, it would be interesting to further investigate
implementations of other concurrent data types and transactional
memory within this operational framework.





\bibliography{references}





\end{document}